\documentclass{ws-procs9x6}

\usepackage{latexsym}
\usepackage{amsmath}
\usepackage{amssymb}
\usepackage{dsfont}
\usepackage{color}  
\usepackage{multirow}
\usepackage{amsmath}
\usepackage{slashbox}
\usepackage{booktabs}

\newcommand{\uk}{\boldsymbol{k}}
\newcommand{\uS}{\boldsymbol{S}}
\newcommand{\sgn}{\text{sgn}}

\begin{document}

\title{MODELING \\THE TRANSVERSE MOMENTUM DEPENDENT\\
PARTON DISTRIBUTIONS}

\author{B. PASQUINI}

\address{Dipartimento di Fisica Nucleare e Teorica, Universit\`a degli Studi di Pavia, and\\
INFN, Sezione di Pavia, I-27100 Pavia, Italy\\
E-mail: Barbara.Pasquini@pv.infn.it}

\author{C. LORC\'E}

\address{Institut f\"ur Kernphysik, Johannes Gutenberg-Universitaet,\\
Mainz, D55099, Germany\\
E-mail: lorce@kph.uni-mainz.de}

\begin{abstract}
We review quark model calculations of the transverse momentum dependent parton distributions (TMDs).
For the T-even TMDs, we discuss the physical origin of model relations
which hold in a large class of quark models.
For the T-odd TMDs we review results in a light-cone constituent quark model (CQM) with the final state interaction effects generated via single-gluon exchange mechanism.  
As phenomenological application, 
we show the good agreement between results in the light-cone CQM and available experimental data for the Collins asymmetry.
\end{abstract}


\bodymatter

\section{Introduction}
\label{sec1}
Parton distributions entering  many hard and exclusive processes 
play a key role to describe the nonperturbative 
structure of hadrons.
The most prominent examples are  the ordinary parton distributions which describe the probability to find in a fast-moving hadron a parton 
with a  certain fraction of the longitudinal momentum of the parent hadron.
However, in addition to the information on the longitudinal momentum distribution,
 a complete three-dimensional picture of the nucleon also requires knowledge of the transverse motion of partons.
A full account of the orbital motion, which is also an important issue to understand the spin structure of the nucleon,  
can be given in terms of transverse-momentum dependent parton 
distributions (TMDs). There are eight leading-twist TMDs.
Two of them, the 
Boer-Mulders and Sivers functions 
are T-odd, i.e. 
they change sign under na\"{\i}ve time reversal, which is defined as 
usual time reversal, but without interchange of initial and final states.
The other six leading-twist TMDs are T-even. 
The latter will be reviewed in Sect.~\ref{sect2}, discussing in particular 
the origin of model relations 
which were found in a large class of 
quark models~\cite{Avakian:2010br,Avakian:2008dz,Pasquini:2008ax,She:2009jq,Efremov:2009ze,Jakob:1997wg,Lorce:2007fa} \!\!\!.
Understanding the physical origin of these relations
provides useful insights for building up realistic quark models.
Indeed, from phenomenological applications to observables one can verify
to which extent the assumptions at the basis of the model relations are 
also satisfied in ``nature''.
The T-odd TMDs will be the subject of Sect.~\ref{sect3}, where we will show predictions  in a light-cone constituent quark model (CQM)~\cite{Pasquini:2010af} 
which was successfully applied 
to describe many nonperturbative properties of the nucleon, like generalized parton distributions~\cite{Boffi:2007yc} \!\!\!, form factors~\cite{Pasquini:2007iz} and distribution amplitudes~\cite{Pasquini:2009ki} \!\!\!.
The same light-cone CQM was also applied 
to describe single-spin asymmetries 
in semi-inclusive deep inelastic (SIDIS) which allow one to extract information
on TMDs~\cite{Boffi:2009sh} \!\!\!.
In Sect.~\ref{sect4} we will show as an example
the results for the Collins asymmetry.
\section{T-even TMDs and quark model relations}
\label{sect2}
In the light-cone gauge $A^+=0$, 
the physical meaning of the correlations encoded in the T-even TMDs becomes 
especially transparent when using  the canonical expansion in terms of light-cone Fock operators for the quark fields. Restricting ourselves to the quark contribution, one finds
\begin{gather}
f^q_1(x,\uk^2_\perp)=\langle P\Lambda|V^q(\tilde k)|P\Lambda\rangle,\label{start}\\
\Lambda\,g^q_{1L}(x,\uk^2_\perp)=\langle P\Lambda|A^q(\tilde k)|P\Lambda\rangle,\\
\frac{\uk_\perp\cdot\uS_\perp}{M}\,g^q_{1T}(x,\uk^2_\perp)=\langle PS_\perp|A^q(\tilde k)|PS_\perp\rangle,\\
\Lambda\,\frac{\uk_\perp}{M}\,h^{\perp q}_{1L}(x,\uk^2_\perp)=\langle P\Lambda|\boldsymbol T^q_\perp(\tilde k)|P\Lambda\rangle,\\
\uS_\perp\,h^q_{1T}(x,\uk^2_\perp)+\frac{\uk_\perp\cdot\uS_\perp}{M}\frac{\uk_\perp}{M}\,h^{\perp q}_{1T}(x,\uk^2_\perp)=\langle PS_\perp|\boldsymbol T^q_\perp(\tilde k)|PS_\perp\rangle.\label{disentangle}
\end{gather}
where $\Lambda$ ($S_\perp$) is the longitudinal (transverse) light-cone polarization
 of the nucleon, and $P$ is the nucleon four-momentum.
In Eqs.~(\ref{start})-(\ref{disentangle}), the quark operators $V^q$, $A^q$, and $\boldsymbol T_\perp^q$ can be written in terms of creation and annihilation operators of quarks with flavor $q$, light-cone helicity $\lambda$ and light-cone momentum $\tilde k=(xP^+,\uk_\perp)$ as
\begin{equation}\label{polarization1}
\begin{split}
V^q(\tilde k)=\sum_\lambda q^\dag_\lambda(\tilde k)q_\lambda&(\tilde k),\quad
A^q(\tilde k)=\sum_\lambda\sgn(\lambda)\,q^\dag_\lambda(\tilde k)q_\lambda(\tilde k),
\\
T^q_R(\tilde k)=&\left[T^q_L(\tilde k)\right]^\dag=2 q^\dag_+(\tilde k)q_-(\tilde k),
\end{split}
\end{equation}
where for a generic two-component vector $\boldsymbol a_\perp$ we have defined $a_{R,L}=a^1\pm ia^2$. The quark operators $V^q$ and $A^q$ have a probabilistic interpretation since they are written just in terms of number operators $N^q_\lambda=q^\dag_\lambda(\tilde k)q_\lambda(\tilde k)$. The operator $\boldsymbol T_\perp^q$ has also a probabilistic interpretation but only when written in terms of transversely polarized operators
\begin{eqnarray}
\uS^q_\perp\cdot\boldsymbol T_\perp^q=\sum_{s_\perp}\sgn(s_\perp)q^\dag_{s_\perp}(\tilde k)q_{s_\perp}(\tilde k).
\end{eqnarray}
Choosing the frame system such that $\hat e_z=\hat P$ and $\hat e_y=\hat k_\perp$,
%$\hat e_1\equiv\hat e_T=\hat e_\perp\times\hat e_3,$  $\hat e_2\equiv\hat e_\perp=\hat k_\perp,$  $\hat e_3\equiv \hat e_L$, 
the T-even TMDs can be classified in terms of
the quark and nucleon light-cone polarizations
as summarized in Table~\ref{table1}.
\begin{center}
\begin{table}[h]
\tbl{T-even TMDs according to quark (columns) and nucleon (rows) 
light-cone polarizations.}
{\label{table1}
\resizebox{0.7\textwidth}{!}{%
\begin{tabular}{c|c|c|c|c}
\toprule
\newline
\backslashbox{nucleon pol}{quark pol}      
        & U
        & L
        & T$_y$
        & T$_x$\cr
\hline\newline
U & $f_1$   &  & & \cr
\newline
L    &       &  $g_{1L}$&  $\frac{k_\perp}{M}h_{1L}^\perp$ &  \cr
\newline
T$_y$   &    &  $\frac{k_\perp}{M}g_{1T}$&$h_{1T}^+$ &\cr
\newline
T$_x$   &  && & $h_{1T}^-$ \\
\bottomrule
\end{tabular}
}
}
\begin{tabnote}
$h_{1T}^{\pm q}=h_{1}\pm\frac{k_\perp^2}{2M} h_{1T}^\perp$.
\end{tabnote}
\end{table}
\end{center}

In QCD all TMDs are independent functions. However, in a large class
of quark models~\cite{Avakian:2010br,Avakian:2008dz,Pasquini:2008ax,She:2009jq,Efremov:2009ze,Jakob:1997wg,Lorce:2007fa} there appear relations among different TMDs.
In particular, for the T-even TMDs one finds $i)$ three linear relations
\begin{gather}
C^q f_1^q+g_{1L}^q-2h_1^q=0,\label{eq:linear1}\\
g^q_{1T}+h^{\perp q}_{1L}=0,\label{eq:linear2}\\
g^q_{1L}-\left[h^q_1+\frac{k^2_\perp}{2M^2}\,h^{\perp q}_{1T}\right]=0,
\label{eq:linear3}
\end{gather}
with $C^q$ in Eq.~(\ref{eq:linear1}) a real number,
% for each quark flavor $q$, 
and
$ii)$ a quadratic relation
\begin{gather}
2 h_1h_{1T}^{\perp\,q}=-(g_{1T}^q)^2.
\label{eq:quadratic}
\end{gather}
Although such model relations hold in quark models based on different descriptions of the quark dynamics, their origin can be traced back to a few common assumptions for treating
the quark and nucleon spin degrees of freedom. 
Indeed, Eqs.~(\ref{eq:linear1})-(\ref{eq:quadratic}) connect the reciprocal polarizations of quark and nucleon, 
and we expect an underlying symmetry under rotation of the polarization degrees of freedom, common to all models. In order to describe the behavior of the polarizations under rotations, it is convenient to work in the basis of canonical 
 (instant form) spin instead of light-cone helicity. This is because rotations in light-cone 
quantization depend non trivially on the interaction,
while they become kinematical operators
within instant-form quantization.
A first common assumption to all the quark models satisfying the relations
 is that there is no 
mutual interaction among the quarks at the interaction vertex.
In this case,
light-cone helicity and canonical spin operators are just related by a Wigner rotation in the polarization space with axis orthogonal to both $\hat e_z$ and $\uk_\perp$ directions
\begin{equation}
q_\lambda(\tilde k)=D^{(1/2)*}_{\lambda s}\,q_s(\tilde k),\qquad 
D^{(1/2)*}_{\lambda s}=\begin{pmatrix}\cos\tfrac{\theta}{2}&\hat k_L\,\sin\tfrac{\theta}{2}\\-\hat k_R\,\sin\tfrac{\theta}{2}&\cos\tfrac{\theta}{2}\end{pmatrix},
\label{eq:matrix}
\end{equation}
where $\lambda$ and $s$ refer to the light-cone helicity and canonical spin, respectively, while
the explicit expression of the angle $\theta$ 
in the different models can be found in Ref.~[\refcite{LorcePasquini}].
Applying the rotation in Eq.~(\ref{eq:matrix}) to the polarization operators in Eq.~(\ref{polarization1}) and the nucleon states, one finds the expressions of the T-even TMDs in the basis of canonical spin given in Table~\ref{table2}.
\begin{table}[h]
\tbl{T-even TMDs according to quark (columns) and nucleon (rows) canonical polarizations.}
{\label{table2}
\resizebox{\textwidth}{!}{%
\begin{tabular}{c|c|c|c|c}
\toprule
\newline
\backslashbox{nucleon pol}{quark pol}
        & U
        & L
        & T$_y$
        & T$_x$\cr
\hline
\newline
U & $f_1$   &  & & \cr
\newline
L    &       &  $\cos\theta \,g_{1L}-\sin\theta \,\frac{k_\perp}{M}h_{1L}^\perp $
&  $\cos\theta \,g_{1L}+\sin\theta \,\frac{k_\perp}{M}h_{1L}^\perp $ &  \cr
\newline
T$_y$   &    &  $\cos\theta\,\frac{k_\perp}{M}g_{1T}-\sin\theta\, h_{1T}^{+}$&
$\sin\theta\,\frac{k_\perp}{M}g_{1T}+\cos\theta\, h_{1T}^{+}$ &\cr
\newline
T$_x$   &  & & & $h_{1T}^-$ \cr
\hline
\end{tabular}
}
}
\end{table}
If we assume rotational symmetry  around one of the coordinate axis,
we find:
\begin{itemize}
\item{}
for the $\hat e_x$ axis: the probability to find a quark 
with spin parallel to the longitudinal direction in a longitudinally polarized nucleon
%and to the nucleon spin 
is equal to the probability to find both quark and nucleon transversely polarized in the $\hat e_y$ direction. As a result, one can derive
from 
Table~\ref{table2} the two linear relations of Eqs.~(\ref{eq:linear2}) and (\ref{eq:linear3});
\item{}
for $\hat e_z$ axis: the momentum distribution is axially symmetric when the canonical spins of the active quark and nucleon are aligned in the same transverse direction.
Using the expressions in Table~\ref{table2}, one finds the quadratic relation of Eq.~(\ref{eq:quadratic}).
\end{itemize}
The cylindrical symmetry around the $\hat e_y$  leads to relations among TMDs which are not independent from the previous ones.
The last linear relation in Eq.~(\ref{eq:linear1}), relating unpolarized and polarized spin amplitudes, holds within more restrictive conditions. It requires spherical symmetry and $SU(6)$ symmetry for the spin-isospin dependence.
We note that all the models discussed in the literature which satisfy 
the relations~(\ref{eq:linear2}) and (\ref{eq:linear3}) among polarized TMDs 
and the quadratic relation~(\ref{eq:quadratic}) satisfy spherical symmetry in the instant-form, which is equivalent to assuming cylindrical symmetry around all the three spatial axis. Note that
 this condition is sufficient but not necessary for the validity of the relations.
\section{T-odd TMDs}
\label{sect3}
\begin{figure}[t]
\begin{center}
\hspace{-2. cm}
\epsfig{file=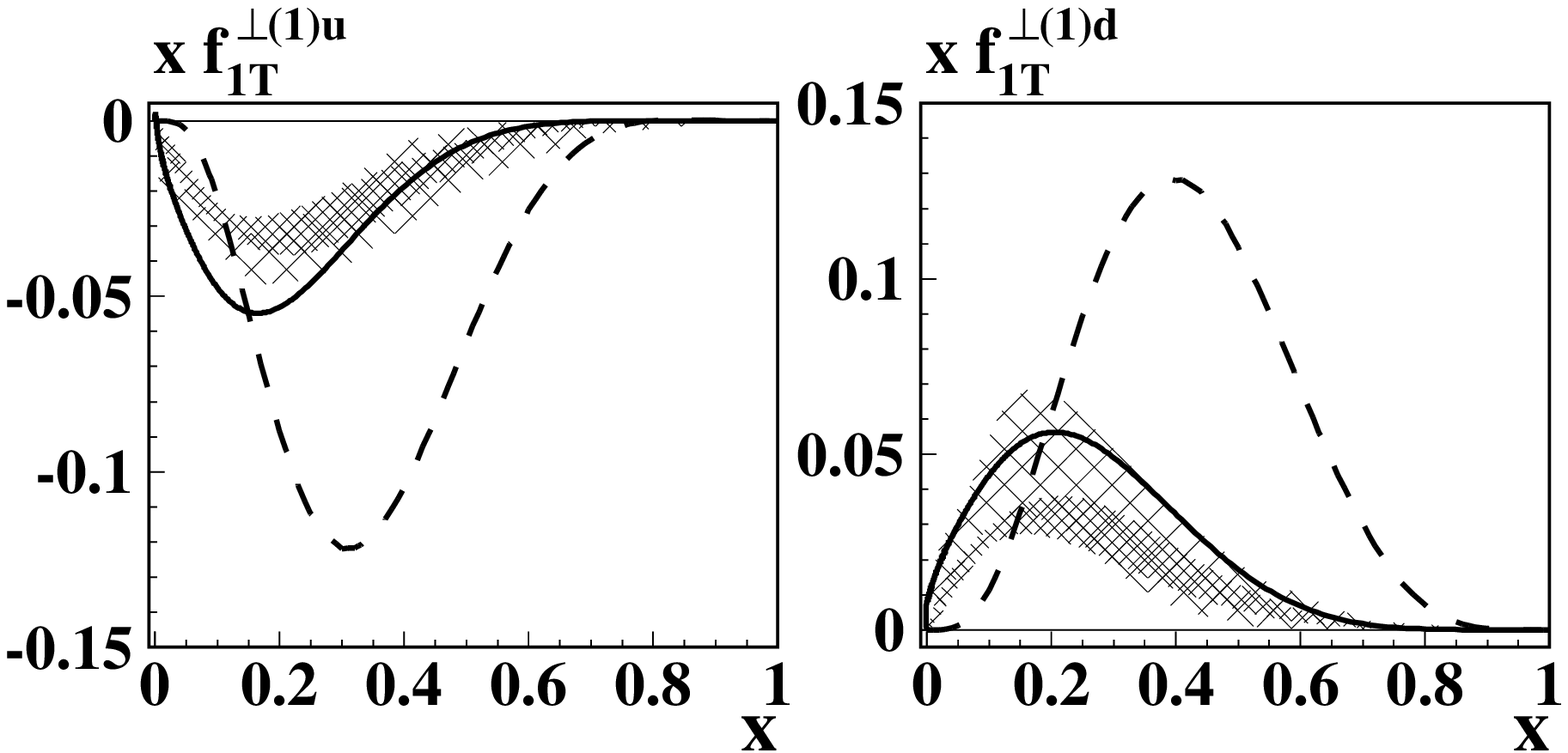,  width=0.38\columnwidth}
\hspace{1. cm}
\epsfig{file=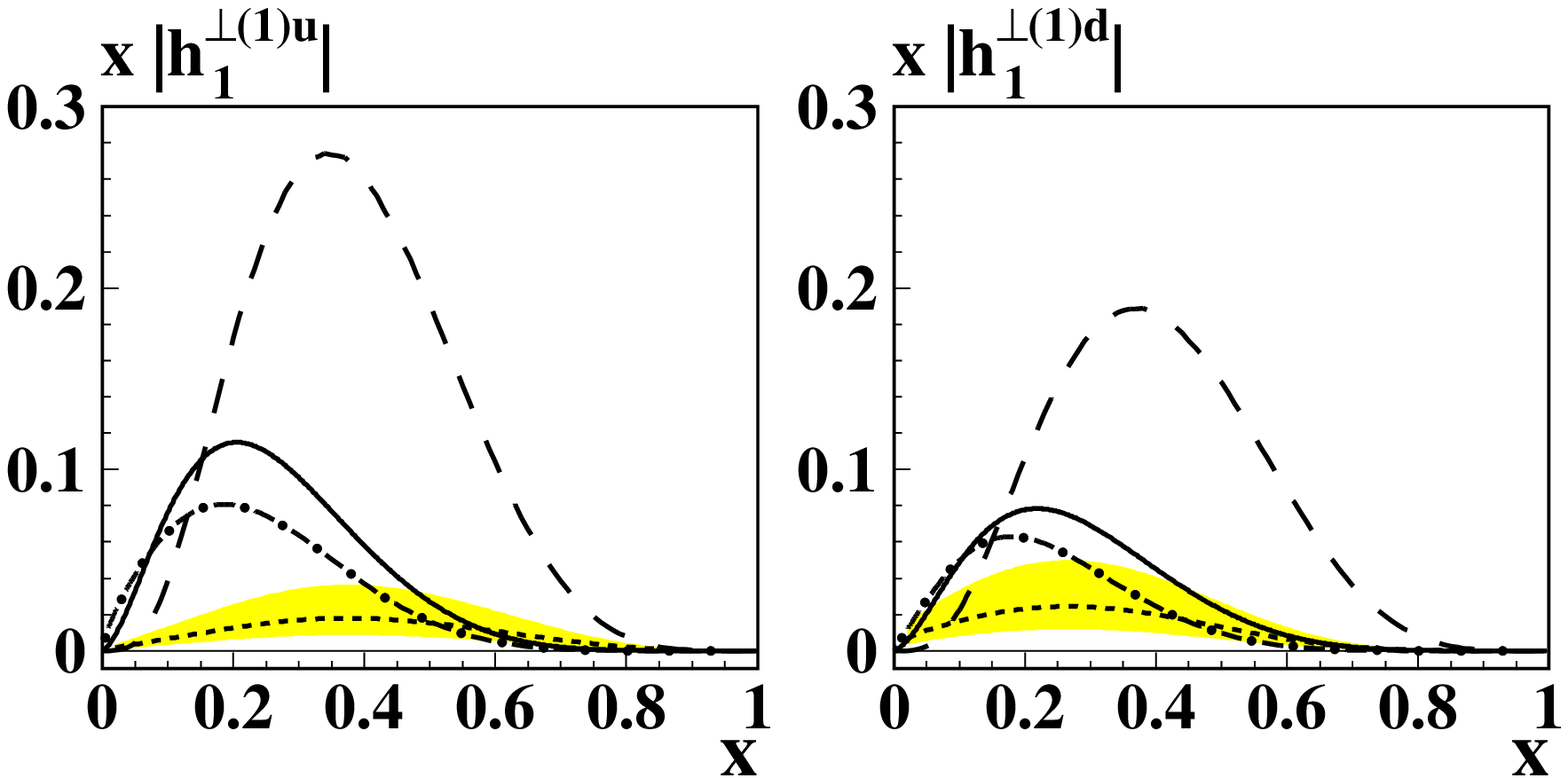,  width=0.38\columnwidth}
\end{center}
\caption{Results for the first transverse-momentum moment of the Sivers and Boer Mulders functions for up and down quarks, as function of $x$. 
See text for the explanation of the different curves.
}
\label{fig2}
\end{figure}

The gauge-link operator entering the quark correlation function which defines the TMDs is crucial to obtain non-zero T-odd quark distributions.
In the light-cone CQM~\cite{Pasquini:2010af} \!\!\!, we consider the expansion of the gauge-link operator
 at leading order, taking into account the one-gluon exchange mechanism between the struck quark and the nucleon spectators described by (real) light-cone wave functions (LCWFs). This approach is complementary to a recent work~\cite{Pasquini:2010af} where the rescattering effects are incorporated in augmented LCWFs containing an imaginary phase.
% which depends on the choice of advanced or retarded boundary conditions for the gauge potential in the light-cone gauge. 
The approximation of truncating the expansion of the gauge-link operator at leading order is common to most of the quark-model calculations in literature~\cite{Bacchetta:2008af,Courtoy:2008dn} \!\!\! .
However, studies beyond the one-gluon exchange approximation where one resums all order contributions
have recently been presented in Ref.~[\refcite{Gamberg:2009uk}].
Following Ref.~[\refcite{Pasquini:2010af}], one can represent the T-odd TMDs in terms of overlap of light-cone amplitudes corresponding to different orbital angular momentum. 
Both T-odd TMDs are obtained from the interference of wave components which differ by one unit of orbital angular momentum.
In particular, we found that the Sivers function for both up and down quarks is dominated by the interference of $S$ and $P$ wave components, while the $P-D$ wave interference terms contribute at most by $20\%$. On the other side, the relative weight of the $P-D$ wave interference terms increases in the case of the Boer-Mulders function, in particular for the down-quark component.
In Fig.~\ref{fig2}
we show the results for the first transverse-momentum moments of the Sivers and Boer-Mulders functions.
The dashed curves correspond to the results at the hadronic scale of the model.
The solid curves are the results evolved to $Q^2=2.5 $ GeV$^2$.
Since the exact evolution equations for the T-odd quark distributions are still under study, we used those evolution equations which seem most promising to simulate the correct evolution.
We evolved the first transverse-momentum moment of the Sivers function by means of the evolution pattern of the unpolarized parton distribution, while for the first transverse-momentum moment of the Boer-Mulders 
we used the evolution pattern of the transversity.
After evolution, the model results are consistent with the available parametrizations. For the Sivers function, the lighter and darker shaded areas are the uncertainty bands due to the statistical error of the parametrizations of Ref.~[\refcite{Anselmino:2008sga}] and Ref.~[\refcite{Collins:2005ie}],
respectively, which refer  to an average  scale of $Q^2=2.5$ GeV$^2$.
For the Boer-Mulders function,
the dashed-dotted curves are the results of the phenomenological parametrization
of Refs.~[\refcite{Barone:2008tn,Barone:2009hw}]
at the average scale of $Q^2=2.4 $ GeV$^2$, and the short-dashed curves
show the results  of
Refs.~[\refcite{Zhang:2008ez,Lu:2009ip}] valid at $Q^2\approx 1$ GeV$^2$, with the shaded area describing the variation ranges allowed by positivity bounds.

We also note that a non trivial constraint in model calculations of the Sivers function is given by the Burkardt sum rule~\cite{Burkardt:2004ur} \!\!\!,
which corresponds to  require that the net (summed over all partons)
transverse momentum due to final-state interaction
is zero.
Our model calculation of the Sivers function
reproduces exactly this sum rule.
\begin{figure}[t]
\centerline{
 \hspace{3.5mm}
\psfig{file=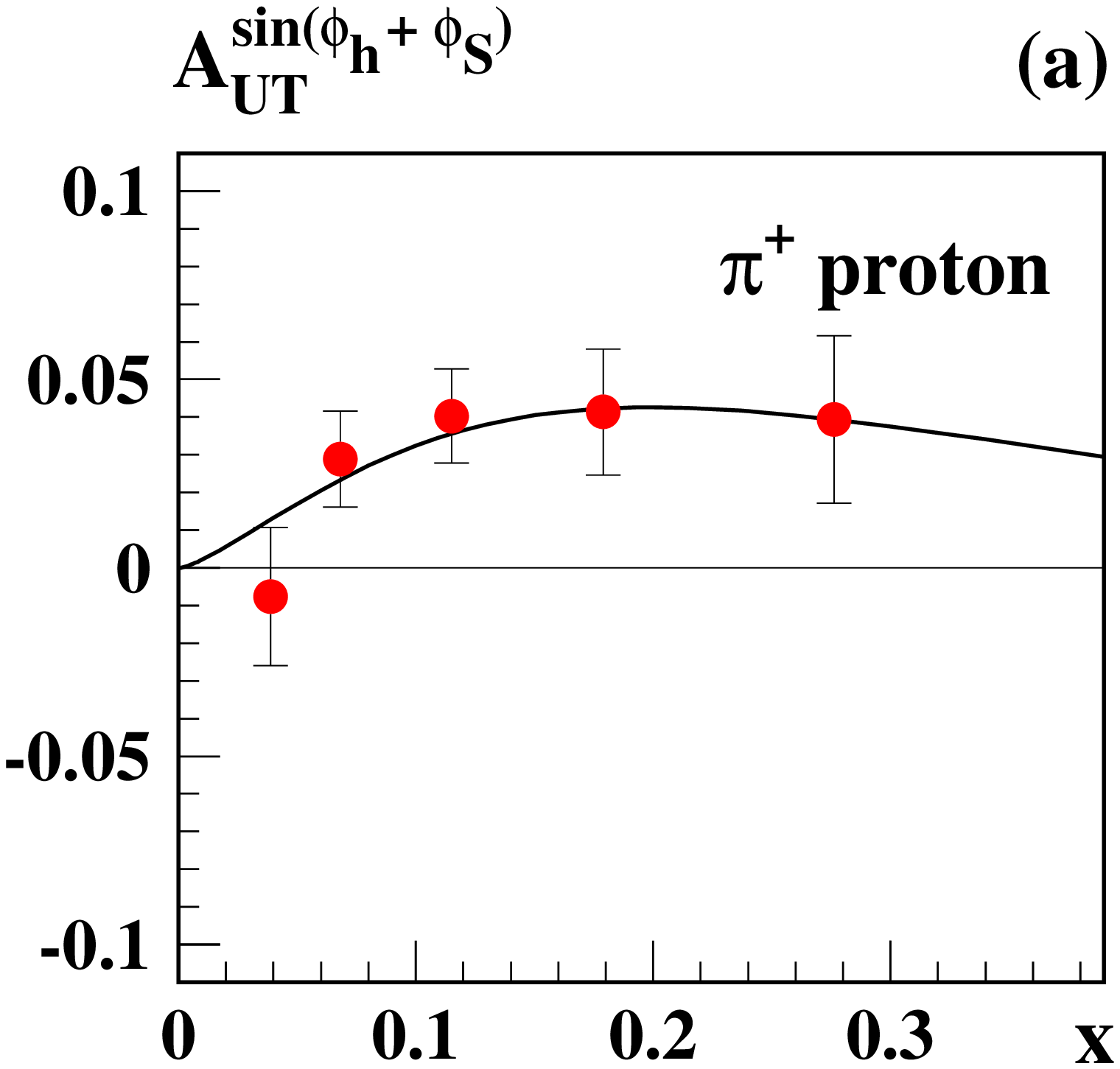, height=3.72 cm}
 \hspace{-11mm}
\psfig{file=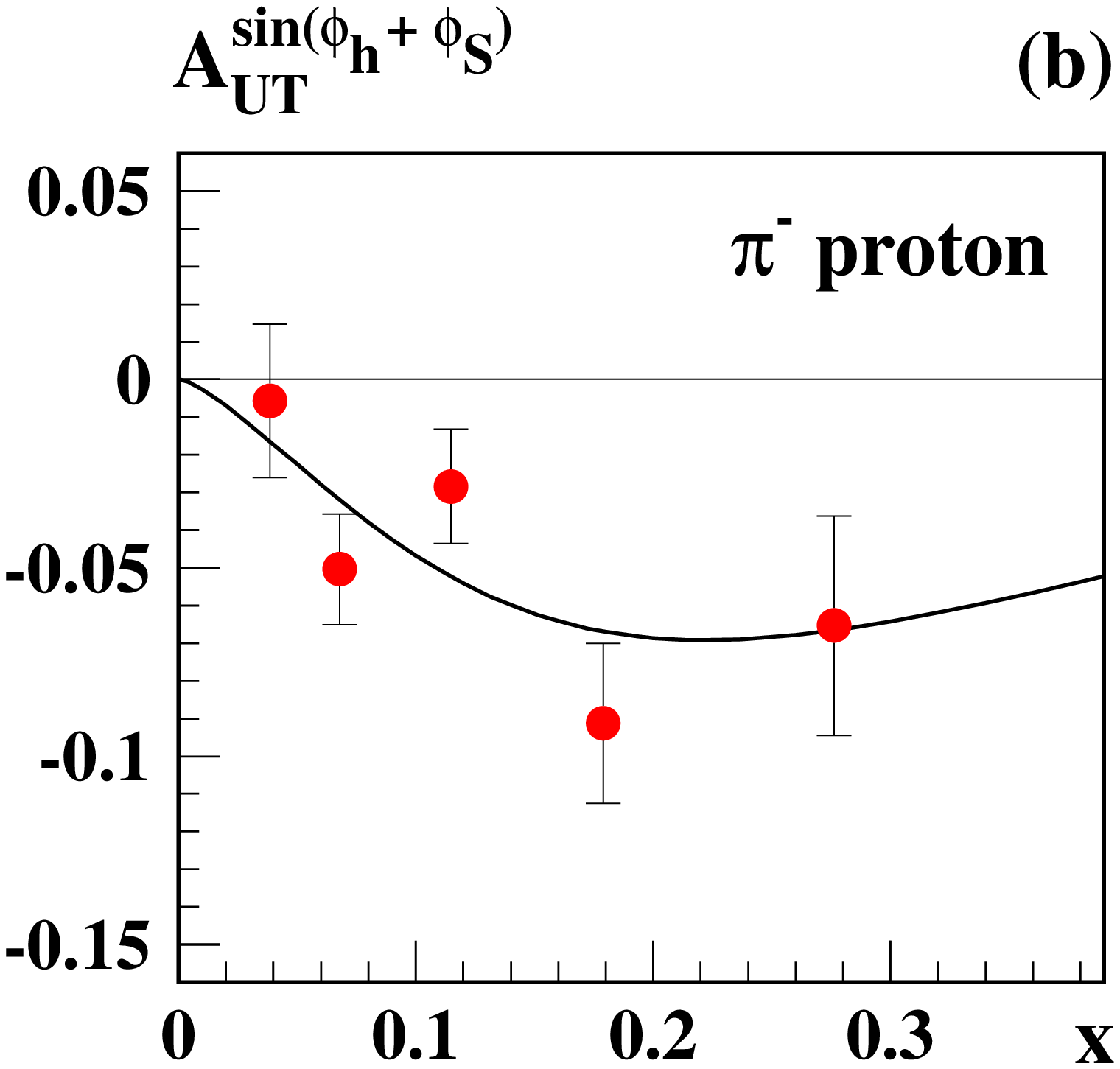, height=3.72 cm}
 \hspace{-12mm}
\psfig{file=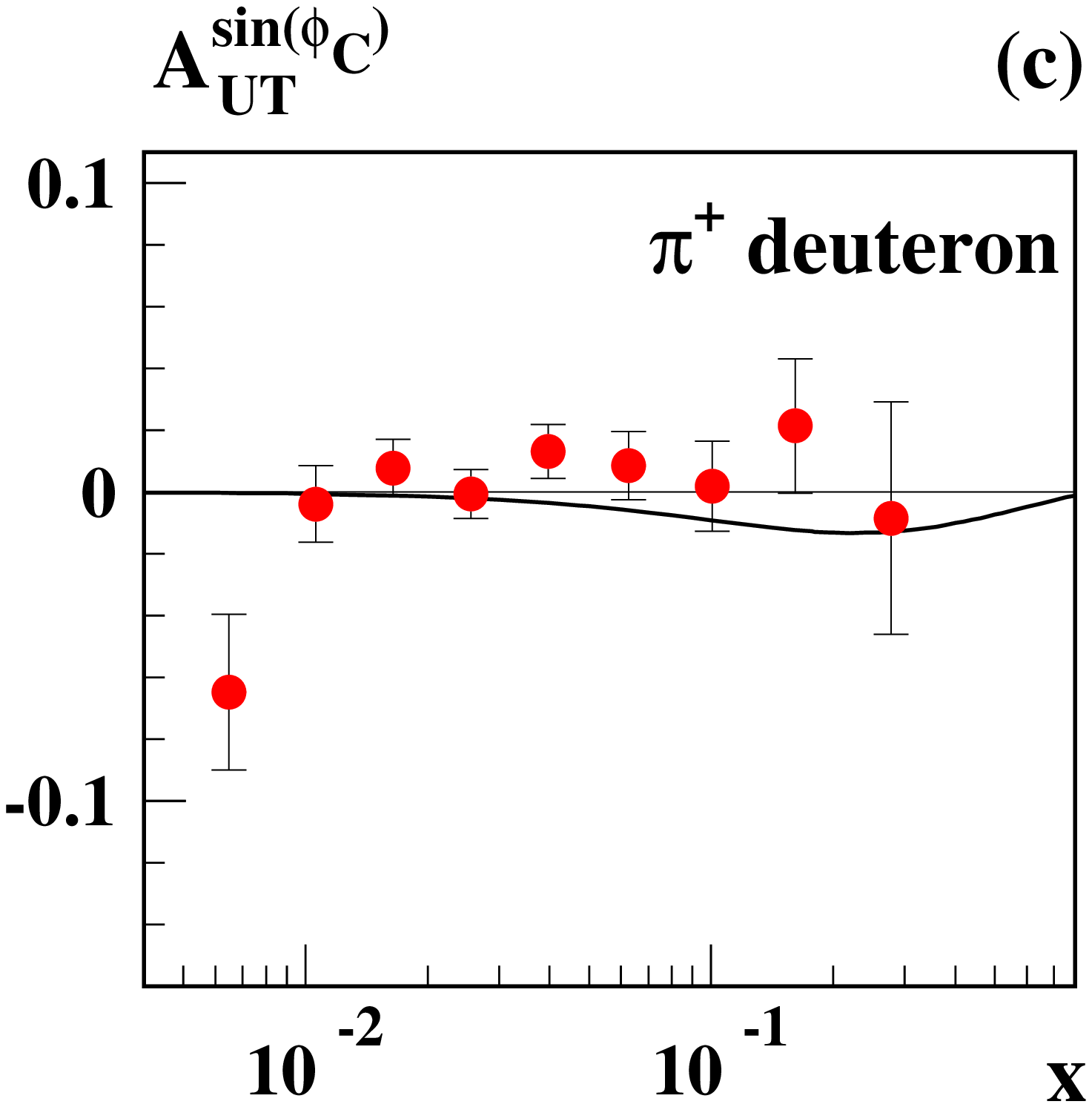, height=3.72 cm}
 \hspace{-11mm}
\psfig{file=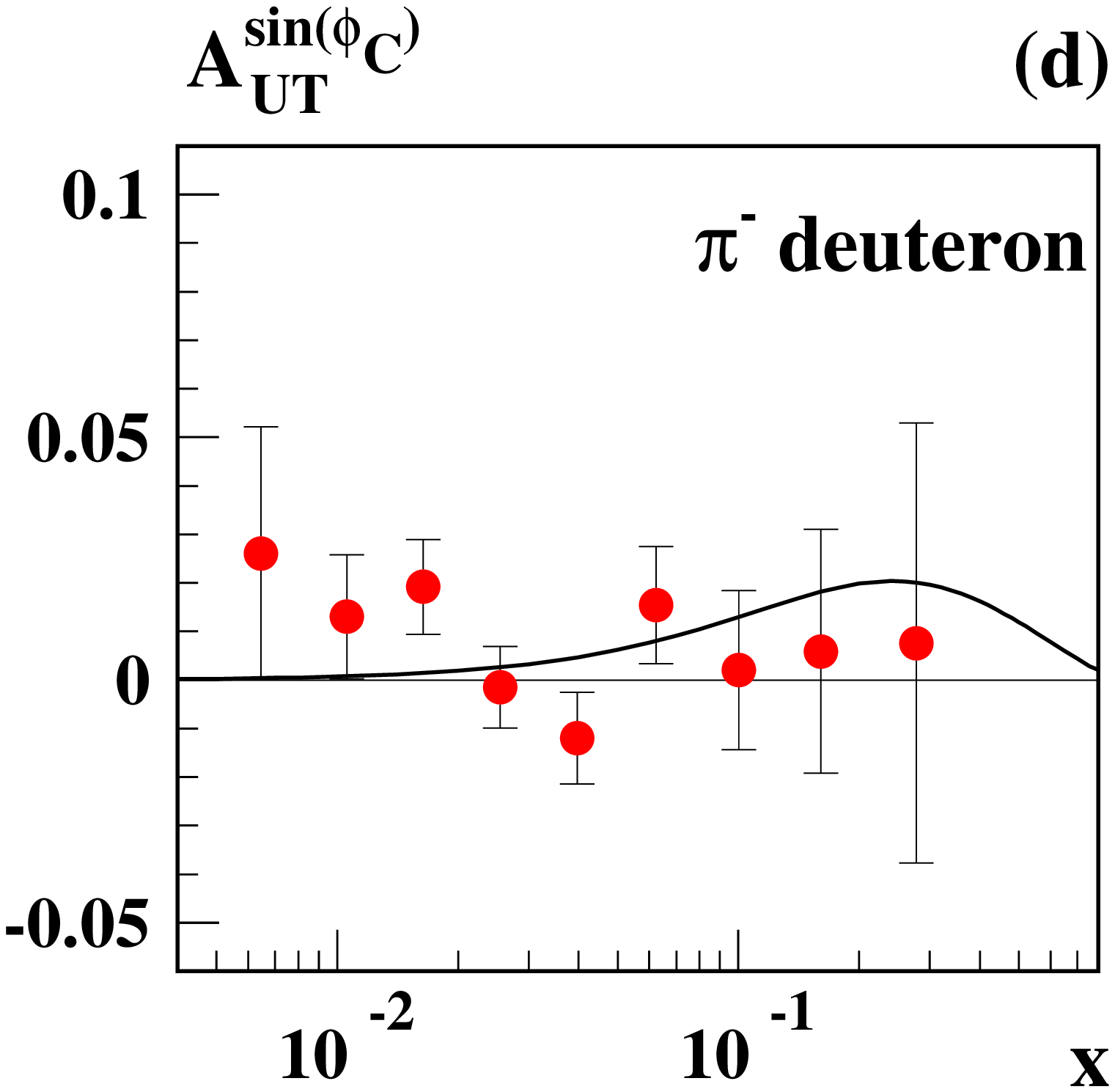, height=3.72 cm}
}
	\caption{\label{fig4}
	The single-spin asymmetry 
        $A_{UT}^{\sin(\phi_h+\phi_S)}\equiv-A_{UT}^{\sin\phi_C}$ in DIS
	production of charged pions 
	off proton and deuterium targets, as function 
        of $x$. The solid curves are obtained using the
        light-cone CQM predictions for $h_1(x,Q^2)$ from
	Refs.~[\protect\refcite{Pasquini:2005dk,Pasquini:2008ax}].
        The proton target data are 
        from HERMES \protect\cite{Diefenthaler:2005gx} \!\!\!,
	the deuterium target data are from COMPASS \protect\cite{Alekseev:2008dn} \!\!\!.}
\end{figure}
\section{Results for the Collins asymmetry}
\label{sect4}

In Ref.~[\refcite{Boffi:2009sh}] the present results for the T-even TMDs 
were applied to estimate azimuthal asymmetries in SIDIS, discussing
the range of applicability of the model, especially with regard to the scale
dependence of the observables and the transverse-momentum dependence of the 
distributions.
Here we review the results  for the Collins asymmetry
 $A_{UT}^{\sin(\phi+\phi_S)}$, due to the Collins fragmentation function and to the chirally-odd TMDs $h_1$.
For the Collins function we use the results extracted in Ref.~[\refcite{Efremov:2006qm}].
In the denominator of the asymmetry we take $f_1$ from Ref.~[\refcite{Gluck:1998xa}]
and the unpolarized fragmentation function from Ref.~[\refcite{Kretzer:2000yf}], both 
at the scale $Q^2=2.5$ GeV$^2$.
\newline
In Fig.~\ref{fig4} the results for the Collins asymmetry
in DIS production of charged pions off proton and deuterium targets are shown 
as function of $x$.
The model results for $h_1$ evolved from the low hadronic scale
of the model to $Q^2=2.5 $ GeV$^2$ ideally describe the HERMES 
data\cite{Diefenthaler:2005gx} for a proton target (panels (a) and (b)).
This is in line with the good agreement between our model predictions\cite{Pasquini:2005dk} and the phenomenological extraction 
of the transversity and the tensor charges\cite{Anselmino:2007fs}.
Our results are also compatible with the COMPASS data\cite{Alekseev:2008dn}
for a deuterium target (panels (c) and (d)) which extend down to much lower values of $x$.

\section*{Acknowledgments}

B.P. acknowledges a fruitful collaboration with S. Boffi, A. Efremov, P. Schweitzer, and F. Yuan with whom some of the results presented in this paper have been obtained.
This work was supported in part  by
the Research Infrastructure Integrating Activity
``Study of Strongly Interacting Matter'' (acronym HadronPhysics2, Grant
Agreement n. 227431) under the Seventh Framework Programme of the
European Community, by the Italian MIUR through the PRIN 
2008EKLACK ``Structure of the nucleon: transverse momentum, transverse 
spin and orbital angular momentum''.
\bibliographystyle{ws-procs9x6}
\bibliography{Pasquini}

\end{document}